\documentclass[a4paper,11pt]{article}
\pdfoutput=1 

\usepackage{jheppub} 

\usepackage[T1]{fontenc} 
\usepackage{comment}
\usepackage{caption}
\usepackage{subcaption}
\usepackage{float}
\usepackage{csquotes}

\graphicspath{{Plots/}}

\def\avg#1{\left\langle#1\right\rangle}
\def\bra#1{\left\langle#1\right|}
\def\ket#1{\left|#1\right\rangle}

\def\abs#1{\left|#1\right|}
\def\kc#1{\left(#1\right)}
\def\kd#1{\left[#1\right]}
\def\ke#1{\left\{#1\right\}}

\def\sgn{\mathrm{sgn}}

\def\nn{\nonumber}

\def\=>{\Rightarrow}
\def\>{\rightarrow}

\def\E{\mathcal{E}}

\DeclareMathOperator\sinc{sinc}

\DeclareMathOperator\Si{Si}
\DeclareMathOperator\Ci{Ci}
\DeclareMathOperator\SFF{SFF}

\title{\boldmath Chaos and integrability in triangular billiards}

\author[a,b,c]{Vijay Balasubramanian}
 \author[d]{Rathindra Nath Das,}
 \author[d]{Johanna Erdmenger,}
 \author[d]{\\and Zhuo-Yu Xian}
 \affiliation[a]{David Rittenhouse Laboratory, University of Pennsylvania,\\209 S. 33rd Street, Philadelphia PA 19104, USA.}
 \affiliation[b]{Theoretische Natuurkunde, Vrije Universiteit Brussel (VUB), and\\International Solvay Institutes, Pleinlaan 2, B-1050 Brussels, Belgium.}
 \affiliation[c]{Santa Fe Institute, 1399 Hyde Park Road,
Santa Fe, NM 87501, USA}
\affiliation[d]{Institute for Theoretical Physics and Astrophysics and \\ W\"urzburg-Dresden Cluster of Excellence ct.qmat \\ Julius-Maximilians-Universit\"at W\"urzburg \\
 Am Hubland, 97074 W\"{u}rzburg, Germany
}

\emailAdd{vijay@physics.upenn.edu, \\das.rathindranath@uni-wuerzburg.de,\\erdmenger@physik.uni-wuerzburg.de \\ zhuo-yu.xian@physik.uni-wuerzburg.de}

\abstract{
We characterize quantum dynamics in triangular billiards in terms of five properties: (1) the level spacing ratio (LSR), (2) spectral complexity (SC), (3) Lanczos coefficient variance, (4) energy eigenstate localisation in the Krylov basis, and (5) dynamical growth of spread complexity.  The billiards we study are classified as integrable, pseudointegrable or non-integrable, depending on their internal angles which determine properties of classical trajectories and associated quantum spectral statistics.  A consistent picture emerges when transitioning from integrable to non-integrable triangles: (1) LSRs increase; (2) spectral complexity growth slows down; (3) Lanczos coefficient variances decrease; (4) energy eigenstates delocalize in the Krylov basis; and  (5) spread complexity increases, displaying a peak prior to a plateau
instead of recurrences.  Pseudo-integrable triangles deviate by a small amount in these charactertistics from non-integrable ones, which in turn approximate models from the Gaussian Orthogonal Ensemble (GOE).  Isosceles pseudointegrable and non-integrable triangles have independent sectors that are symmetric and antisymmetric under a reflection symmetry.  These sectors separately reproduce characteristics of the GOE, even though
the combined system approximates characteristics expected from integrable theories with Poisson distributed spectra.
}

\begin{document} 
\maketitle
\flushbottom

\section{Introduction}

Chaos and integrability remain poorly understood in quantum systems.  They are primarily characterized via spectral statistics:  maximally chaotic systems are believed to contain level spacing correlations matching those of random matrix ensembles, while fully integrable systems have uncorrelated Poisson spectra \cite{Bohigas:1983er,Berry:1985semiclassical,Muller:2004semiclassical, Haake2010,Dyson1972,Guhr:1997ve,Lozej2022}.  However, many quantum systems show behaviour somewhere between integrability and maximal chaos, and we do not understand how to precisely quantify and characterize them. Some information about the early dynamics is available in out-of-time-ordered correlators (OTOCs), which show Lyapunov-like growth for general chaotic systems, with a bound on this growth controlled by unitarity \cite{Maldacena_2016,Garc_a_Mata_2023}. Here, we apply a new tool in the study of quantum dynamics --  spread complexity \cite{Balasubramanian:2022tpr}
-- to characterize differences between integrable,  chaotic, and an intermediate ``pseudo-integrable'' quantum dynamics \cite{RICHENS1981495, Jain_2017}.

To this end, we study systems in which these three types of dynamics are accessed by changing a few parameters: triangular quantum billiards \cite{RICHENS1981495}.  In the classical billiards, a free particle bounces elastically off enclosure walls;  quantum mechanically \cite{sep-ergodic-hierarchy, Ott2002, PhysRevResearch.5.033126,Jha2014, PhysRevE.50.2355, Samajdar_2014, Samajdar_2018, Harish_1996, Jain_2017} the dynamics are determined by the two-dimensional Schr\"odinger equation with vanishing potential inside a triangular boundary where the potential diverges.  Triangular billiards are not strongly chaotic -- their classical Lyapunov exponent is zero because initially close trajectories separate linearly over time, and the Kolmogorov-Sinai entropy vanishes \cite{Jain_2017, PhysRevA.42.3170, MOUDGALYA201582,Lozej2022}. Nevertheless, these systems can exhibit ergodic and mixing behaviour \cite{PhysRevE.109.014224, N_Chernov_1998}. The degree of integrability/chaos is  controlled by the internal angles. For example, the equilateral triangle is classically integrable -- the motion can be solved by quadratures in terms of constants of motion.  Generic triangles with rational angles are pseudo-integrable --  beams of trajectories split at triangle vertices so that no global coordinate transformation produces momenta that are constants of motion \cite{RICHENS1981495}, while the quantum spectra show level repulsion characteristic of chaos, despite an exponential tail in the level spacing distribution characteristic of integrability \cite{Lozej:2024tpt}. Meanwhile, triangles with irrational angles show mixing dynamics and have spectral statistics resembling matrices drawn from the Gaussian Orthogonal Ensemble (GOE) \cite{Lozej2022}. The symmetry of the triangle, e.g., whether it is isosceles or right-angled, also affects dynamics.  Sec.~\ref{sec:billiard-review} reviews the theory of triangular billiards and tabulates the integrable, pseudo-integrable and non-integrable triangles that we study.

In general, chaos leads to spectral correlations, while integrable theories have uncorrelated spectra. The authors of \cite{Iliesiu:2021ari} proposed an integral transform of spectral density correlations as a notion of  ``spectral complexity'' which has been used to study the stadium billard  \cite{camargo2023spectral}. Thus, in Sec.~\ref{sec:spectral_stat}, we discuss spectral properties -- level spacing ratio (LSR), spectral form factor (SFF), and spectral complexity (SC) \cite{Iliesiu:2021ari} -- and compare with analytical results for Poisson and GOE spectra.  Generic pseudo-integrable and non-integrable triangles have level spacings and LSRs resembling the GOE, consistently with \cite{Lozej2022, Lozej:2024tpt}, despite deviations due to numerical Hilbert space truncation and scarring.  Symmetry causes larger deviations  -- isosceles pseudo-integrable and non-integrable triangles display Poisson-like LSRs, because eigenvalues of states even or odd under the triangles' reflection symmetry are uncorrelated, although these sectors separately display GOE-like statistics. Right triangles show harder to characterise deviations from GOE behaviour. Finally, integrable triangles have lower-than-Poisson LSRs partly because of spectral degeneracies.   We show analytically that late time spectral complexity grows linearly for non-degenerate Poisson spectra while  GOE spectra show logarithmic growth, with eventual saturation if the Hilbert space is finite-dimensional.  Meanwhile, energy level degeneracies lead to quadratic spectral complexity growth. Consistently, we find that the late time SC of integrable triangles, which have substantial spectral degeneracies, shows quadratic growth, while generic pseudo-integrable and non-integrable triangles exhibit logarithmic behaviour with eventual saturation. Isosceles pseudo-integrable and non-integrable triangles show linear SC growth close to the  Poisson result, but the symmetric and anti-symmetric sectors deviate strongly, reflecting both the overall Poisson-like spectral character and  GOE-like structure of individual sectors.

Chaos should scramble information \cite{Sekino:2008scramblers} and spread operators \cite{Roberts:2014localized,Roberts:2018operator} faster than integrable dynamics. The authors of \cite{Parker:2018a} proposed a measure of operator spread showing universal behaviour at early times \cite{Parker:2018a,Kar:2021nbm,Caputa:2021sib,Dymarsky:2021bjq,Barbon:2019on,Avdoshkin:2019euclidean,Jian:2020qpp,Nandy:2024htc} and carrying signatures of late time chaos \cite{Rabinovici:2020operator,Dymarsky:2019quantum,Rabinovici:2021qqt,Rabinovici:2022beu}.  We also expect chaos will be more effective than integrable dynamics at transporting generic initial states throughout the Hilbert space. The authors of \cite{Balasubramanian:2022tpr} proposed ``spread complexity'' to quantify this transport. Briefly, the idea is to measure the breadth of support of the wavefunction in a unique minimizing basis known as the Krylov basis.  In this basis, the Hamiltonian is tridiagonalized, and, for Random Matrix Theories which model maximal chaos, the density of states analytically determines the tridiagonal coefficients, also called Lanczos coefficients \cite{Balasubramanian:2022dnj,Balasubramanian:2023kwd}. The temporal variation of spread complexity quantifies information spreading and chaos in quantum systems \cite{Erdmenger:2023wjg,Balasubramanian:2023kwd}.

Thus, in Sec.~\ref{sec:krylov_results}, we compare spread complexities of integrable, pseudo-integrable, and chaotic triangular billiards starting from initial states uniformly spread across the energy eigenbasis.  We show that the variance of the Lanczos coefficients, i.e., non-zero entries of the tri-diagonalized Hamiltonian, is highest for integrable triangles, followed by isosceles pseudo-integrable and non-integrable triangles, and lowest for the generic triangles. The Lanczos coefficient variances turn out to be inversely related to the LSR, thus relating spread complexity to the spectral properties in Sec.~\ref{sec:spectral_stat}.  In the Krylov basis, every Hamiltonian describes a one-dimensional chain with hopping probabilities set by the Lanczos coefficients (see \cite{Balasubramanian:2022tpr} for a discussion in the context of spread complexity). Hence Lanczos coefficient variances effectively correspond to disorder in the Krylov chain transition elements.  The large Lanczos coefficient fluctuations in integrable and symmetric billiards have a consequence -- localisation of the energy basis in the Krylov space as measured by the inverse participation ratio.  The localisation is highest for integrable triangles, followed by pseudo-integrable and non-integrable isosceles triangles, pseudo-integrable and non-integrable right triangles, and generic triangles. For integrable triangles,  spread complexity grows, saturates, and then displays recurrences.  For the other triangles, the spread complexity reaches a plateau at late times after initial growth and fluctuations. Isosceles triangles exhibit slower initial growth, and do not show the complexity peak preceding a slope down to saturation that is characteristic for chaotic theories \cite{Balasubramanian:2022tpr}, and is associated with classical Lyapunov exponents \cite{Hashimoto:2023swv}.  Restricting to the symmetric or anti-symmetric sectors in isosceles triangles reveals chaotic behaviour similar to generic triangles.  This distinction is amplified in the higher moments of spread complexity.

We conclude in Sec.~\ref{sec:Discussion} by discussing future directions.

\paragraph{Note added.} While this article was being prepared, \cite{camargo2024spread} appeared, using Krylov and spectral complexity to explore integrable and chaotic XXZ spin chains.

\section{Setup of the billiards}\label{sec:billiard-review}

In a classical billiard, a frictionless particle travels within a bounded region, undergoing elastic collision at the edges. In a quantum billiard, the wave function of a free quantum particle evolves within a two-dimensional box with an infinite barrier at the boundaries. 

We numerically solve the Schr\"{o}dinger equation for a quantum billiard 
\begin{align}
    -\nabla^2\psi_j=\mathcal E_j\psi_j,\quad j=0,1,2,\cdots
\end{align}
with Dirichlet boundary conditions $\psi_j(x_\partial)=0$ at points  $x_\partial$  on the boundary of a triangluar region. Without loss of generality,  we take the particle mass to be $m=\frac{1}{2}$, set $\hbar=1$, and constrain every triangle to unit area.  For each billiard we calculate the lowest $D_0=1000$ energy levels $\ke{\E_j}_{j=0}^{D_0-1}$, and also separately for the symmetric and antisymmetric sectors of isosceles triangles, which are isolated by a reflection symmetry.

\subsection{Choice of triangles: topology and ergodicity of classical motion \label{subsec:triangle_topology}}

The shape of a triangular billiard determines the phase space of a classical particle moving within it, and whether the trajectories realize integrable, so-called ``pseudo-integrable'', or chaotic dynamics \cite{sep-ergodic-hierarchy, Ott2002,Jain_2017, PhysRevResearch.5.033126}. The phase space for particle motion within a two-dimensional billiard can be represented by coordinates $q = (x, y)$ and motion direction, $\theta$, because the magnitude of the momentum is conserved. When the trajectory encounters a boundary, the particle reflects \cite{Zemlyakov1975, RICHENS1981495}. It is convenient to represent this process by replicating the enclosure, sewing the replica at the corresponding boundary, and continuing the trajectory into the replica. If the enclosed angles of the triangle are {\it rational} fractions of $\pi$, the particle's trajectory traverses a finite number of such sheets \cite{Zemlyakov1975, RICHENS1981495, Jain_2017} before recurring. The continuous surface produced by mirroring and attaching the traversed triangles is called the {\it translation surface} (alternatively the {\it invariant surface}  \cite{RICHENS1981495}). Since the number of traversed sheets and the enclosure size are finite for rational billiards, the translation surface will be compact by construction.

For a triangle with  rational internal angles $\{\frac{p_1}{q_1}, \frac{p_2}{q_2}, \frac{p_3}{q_3}\}$, with $p_i,q_i \in Z^+$ and relatively prime, the genus of  the translation surface is \cite{Zemlyakov1975}:
\begin{equation}
g = 1 + \frac{\mathcal{N}}{2}\sum_i\left(\frac{p_i-1}{q_i}\right)
\end{equation}
Here, $\mathcal{N}$ is the lowest common multiple of the $q_i$, and denotes the number of sheets necessary to construct the translation surface.  Triangles compactifying to genus one have classically integrable dynamics according Arnold's criterion --  there are $n$ constants of motion for $n$ degrees of freedom, so  dynamics restricts to an $n$-torus and can be solved by quadratures \cite{Arnold1989,Arnold1989,RICHENS1981495}.  Triangles compactifying to a finite genus greater than one are said to be pseudo-integrable because, despite conservation of momentum in the replicated and sewn billiard, and contrary to the expectation from Arnold \cite{Arnold1989}, the phase space dynamics is restricted a {\it higher} genus two-dimensional surface \cite{RICHENS1981495}.  This non-integrability arises because beams of trajectories intersecting some vertices of the triangle become separated into different handles of the replicated geometry \cite{RICHENS1981495}. Finally, if any internal angle is irrational, the translation surface has an infinite genus and is homeomorphic to the Loch Ness Monster,  a non-compact, orientable, infinite genus surface with one end (a single-component ideal boundary) \cite{Valdez2009}.  Motion in these irrational triangles is not integrable and shows signatures of chaos like level repulsion \cite{Lozej2022} even though nearby trajectories separate linearly (vanishing Lyapunov exponent).   We consider triangles from all three classes (Table~\ref{tab:trig_list}).

\begin{table}[hbtp]
\centering
\renewcommand{\arraystretch}{1.2}
\begin{tabular}{|c|c|c|c|c|}
\hline
 & g & Isosceles  & Right & General \\
\hline
I & $1$ & $\{\frac{1}{3},\frac{1}{3},\frac{1}{3}\}_1$, $\{\frac{1}{2},\frac{1}{4},\frac{1}{4}\}_2$ & $\{\frac{1}{2},\frac{1}{3},\frac{1}{6}\}_3$ &  \textemdash \\
\hline
 & $2$ & $\{\frac{2}{5},\frac{2}{5},\frac{1}{5}\}_4$, $\{\frac{1}{5},\frac{1}{5},\frac{3}{5}\}_5$ & $\{\frac{3}{8},\frac{1}{8},\frac{1}{2}\}_6$, $\{\frac{2}{5},\frac{1}{10},\frac{1}{2}\}_7$ &   \textemdash \\
 & $3$ & $\{\frac{3}{7},\frac{3}{7},\frac{1}{7}\}_8$ & $\{\frac{5}{12},\frac{1}{12},\frac{1}{2}\}_9$ & $\{\frac{1}{3},\frac{1}{4},\frac{5}{12}\}_{10}$\\
PI  & $4$ & $\{\frac{1}{9},\frac{4}{9},\frac{4}{9}\}_{11}$ & $\{\frac{4}{9},\frac{1}{18},\frac{1}{2}\}_{12}$ & $\{\frac{1}{4},\frac{1}{6},\frac{7}{12}\}_{13}$ \\
 & $5$ & $\{\frac{5}{11},\frac{5}{11},\frac{1}{11}\}_{14}$ & $\{\frac{9}{20},\frac{1}{20},\frac{1}{2}\}_{15}$ & $\{\frac{4}{15},\frac{1}{15},\frac{2}{3}\}_{16}$\\
  & $6$ & $\{\frac{6}{13},\frac{6}{13},\frac{1}{13}\}_{17}$ & $\{\frac{11}{24},\frac{1}{24},\frac{1}{2}\}_{18}$ & $\{\frac{5}{16},\frac{7}{16},\frac{1}{4}\}_{19}$\\
\hline
NI  & $\infty$ & \textemdash & $\{\frac{(-1 + \sqrt{5})}{8},\frac{(5 - \sqrt{5})}{8},\frac{1}{2}\}_{20}$,  & $\{\frac{(3 - \sqrt{5})}{5},\frac{3}{5},\frac{(-1 + \sqrt{5})}{5}\}_{22}$, \\
&&& $\{\frac{(6 - \sqrt{5})}{8},\frac{(-2 + \sqrt{5})}{8},\frac{1}{2}\}_{21}$&$\{\frac{(-2 + \sqrt{5})}{8},\frac{3}{4},\frac{(4 - \sqrt{5})}{8}\}_{23}$\\
\hline
NI&$\infty$&  $\{\frac{1}{4\sqrt{2}},\frac{1}{4\sqrt{2}} ,\frac{(-4 + \sqrt{2})}{4}\}_{24}$ &\textemdash&  $\{\frac{(1 + \sqrt{2})(3 - \sqrt{5})}{4(3 + \sqrt{2})},\frac{(1 + \sqrt{5})}{4},\frac{(3 - \sqrt{5})}{2(3 + \sqrt{2})}\}_{26}$\\
&& $\{\frac{1}{2\sqrt{2}},\frac{1}{2\sqrt{2}},\frac{\sqrt{2}-1}{\sqrt{2}}\}_{25}$&& $\{\frac{(-1 + \sqrt{5})}{8},\frac{(5 - \sqrt{5})}{16},\frac{(13 - \sqrt{5})}{16}\}_{27}$\\
\hline
\end{tabular}
\caption{ 27
triangles are studied in this paper. We denote each triangle as $\{\frac{p_1}{q_1},\frac{p_2}{q_2},\frac{p_3}{q_3}\}_\kappa$ where the internal angles are $p_i \pi/q_i$ and, for convenience, each triangle is indexed by an integer $\kappa$.
I $\equiv$ Integrable triangle, PI $\equiv$ Pseudo-integrable triangle, NI $\equiv$ Nonintegrable triangle, g $\equiv$ Genus of the  translation surface.  All three known integrable triangles have been included.   $\{\frac{1}{2}, \frac{1}{4},\frac{1}{4} \}_2$ is both isosceles and right angled, and triangles with genus one or two translation surfaces are exclusively  isosceles or right-angled.  The first non-integrable group has one rational angle, while the second group has no rational angles.
}
\label{tab:trig_list}
\end{table}

\begin{figure}[H]
    \centering
    \includegraphics[width=0.9\textwidth]{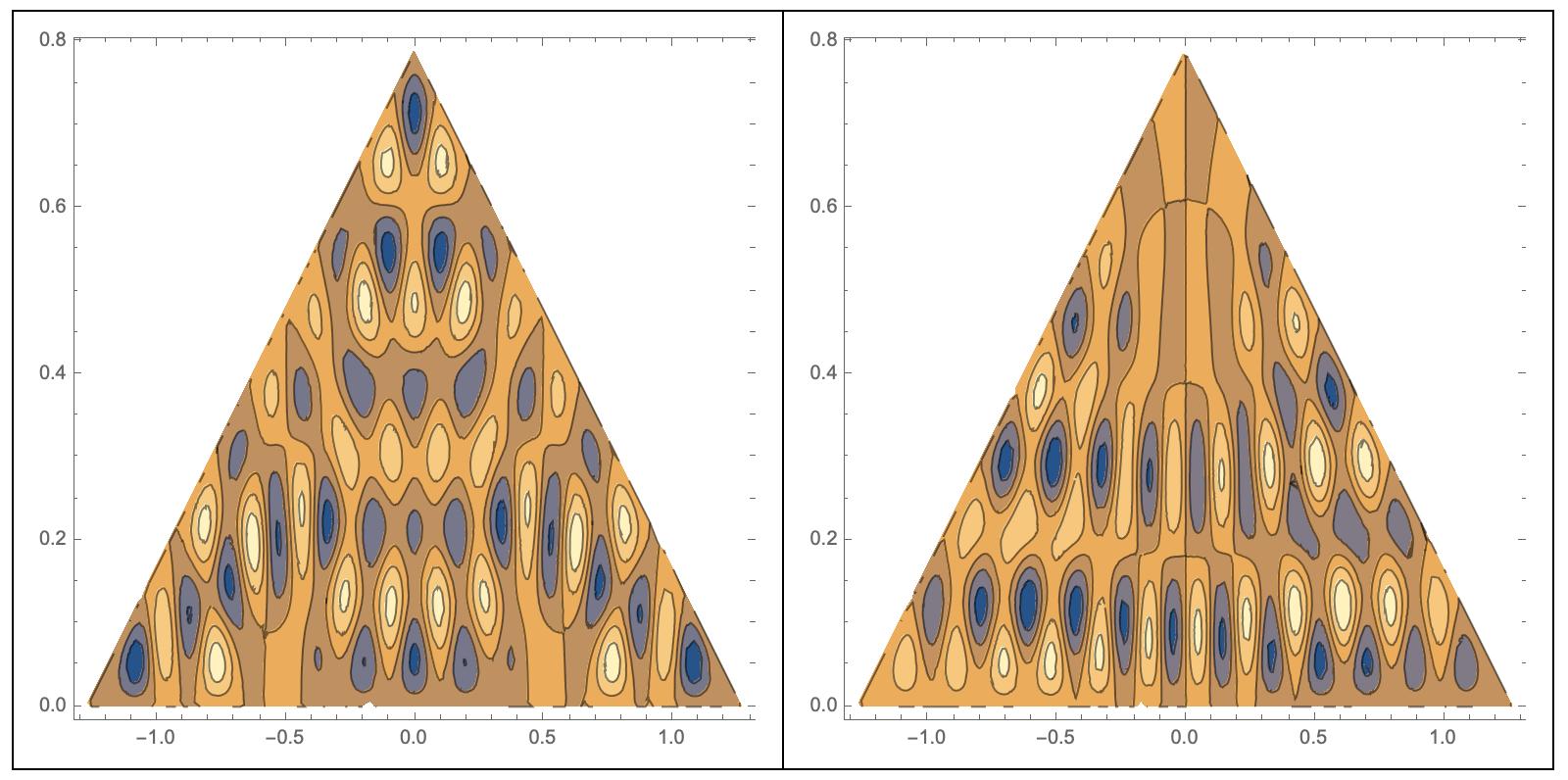}
    \caption{92nd (left) and 95th (right) energy eigenstate of $\kappa$=24 (irrational isosceles) triangle. The left (right) panel shows a symmetric (an anti-symmetric) eigenstates with respect to the reflection axis.}
    \label{fig:symm_antisymm_eigenstate}
\end{figure}

The dynamics of a system can be affected by the presence of isolated symmetry sectors, or by fine-tuning parameters, for example, by selecting one angle of an otherwise irrational triangle to be 90 degrees \cite{Lozej2022}.  To test the effects of symmetry, we consider isosceles triangles (first column in Table~\ref{tab:trig_list}), which have a reflection symmetry across the diagonal, so that the Hamiltonian commutes with the reflection operator. Consequently, all energy eigenstates of isosceles triangles are either symmetric or anti-symmetric with respect to this reflection axis (Fig.~\ref{fig:symm_antisymm_eigenstate}).  The spectral statistics and ergodic properties of the full system can differ from those in the symmetric or anti-symmetric sectors.
To test the effects of fine-tuning, we consider right triangles (second column in Table~\ref{tab:trig_list}).  Right triangular irrational billiards are known to show weak, but not strong, mixing \cite{Artuso1997, Wang2014, Huang2017}, although the dynamics is ergodic \cite{Lozej2022}.

\section{Spectral statistics of triangular billiards}\label{sec:spectral_stat}

The degree of quantum chaos is controlled by short-range correlations and fluctuations in the energy eigenvalues.  We want to compare these spectral statistics between the triangular billiards and chaotic, random matrix theories (RMTs), independently of the coarse-grained spectral density profile, which depends on non-universal characteristics like the overall area of the billiard.  To do so, a standard procedure is to apply an ``unfolding'' process \cite{Haake2010} that uniformizes the 
spectral density so that the local statistics of different triangles can be compared.  Given a density of states $\bar{\rho}(E)$ coarse-grained at some scale that is sufficiently larger than the typical energy gap, such a uniformisation can be realized by mapping
\begin{equation}
    e_j = \bar N(\E_j),\quad \forall j,
\end{equation}
where $\bar N(\E_j) = \int_{-\infty}^{\E_j} dE \, \bar{\rho}(E)$ is the cumulative number of energy eigenvalues below $\E_j$.
This procedure renders the spectrum $\ke{e_j}$ macroscopically uniform while maintaining local fluctuations.  However, unfolding effectively changes the Hamiltonian and will lead to different results for dynamical properties like the spread of the wavefunction.  Hence, we  follow an alternate procedure \cite{Bohigas:1983er,Miltenburg1994,Lozej2022,yczkowski1992ClassicalAQ,Cheon:levelpseudointegrable,Bogomolny2021,Gorin_2001,barrierbilliards}.

First,  note that for triangle billiards, the cumulative number of energy levels below $\E$ asymptotes at large $\E$, after coarse-graining over a suitable scale larger than the typical gap, to the generalized Weyl formula \cite{Baltes1977,Miltenburg1994,Lozej2022}
\begin{equation}\label{eq:WeylFormula}
    \bar N(\E) = \frac{A \E-L\sqrt{\E}}{4\pi} + \frac{\pi}{24}\kc{\frac1\alpha+\frac1\beta+\frac1\gamma-\frac1\pi},
\end{equation}
with $\alpha$, $\beta$, $\gamma$ the three internal angles, $A$ the area of the triangle (set to $1$ in our analyses), and $L=A\frac{\sqrt{2} (\sin \alpha +\sin \beta +\sin \gamma )}{\sqrt{\sin \alpha  \sin \beta  \sin \gamma }}$ its perimeter.  
When $\E\gg (L/A)^2$, the cumulative count given by this ``spectral staircase'' formula is approximately linear in $\E$: $\bar N(\E) \approx A \E / 4\pi$. So, for sufficiently large energies, the coarse grained spectral density in any triangle is already uniform. Thus, in most  analyses, we  examine the truncated  spectrum:
\begin{equation}
    \ke{\E_j}_{j=M}^{D_0-1} \quad  \text{with} \quad \E_{M}\gg (L/A)^2  \quad \text{and} \quad D \equiv D_0  - M \, .
\label{eq:Truncate}
\end{equation}
By doing so, we avoid the need for unfolding. In our numerical studies we examine the first $D_0 = 1000$ levels and find that for all the triangles we consider the Weyl formula linearizes for $j\gtrsim 50$ so that we can take $M=50$. Finally, we  shift and rescale the truncated spectrum $\ke{\E_j}_{j=M}^{D_0-1}$ 
so that the mean energy is $0$, while the spread of energies $E_D-E_0$ is $1$:
\begin{equation}
E_{j}=\frac{\E_{j+M}-\bar{\E}}{\E_{D_0-1}-\E_M},\quad 
\bar \E=\frac1D\sum_{j=M}^{D_0-1} \E_j.
\label{eq:energyrescaling}
\end{equation}
This normalisation rescales the truncated spectrum to always occupy the same range so that we can consistently compare different triangles.

\subsection{Level Spacing Ratio}
The Level Spacing Ratio (LSR), which measures the ratio of the smaller and larger energy gaps around a level, differentiates spectral fluctuations for Poisson or random matrix statistics, {\it without} unfolding \cite{Atas:2013gvn}.  Thus we  use the  full spectrum up to the bound $D_0$ in \eqref{eq:Truncate} to examine LSRs for the triangle billiards, without  the lower cutoff $M$.  Denote by $p(s)$ the histogram of level spacings $\ke{s_j|j=1,2,\cdots,D-1}$ where $s_j=\E_{j+1}-\E_j$ is the difference between consecutive energy levels $s_j=\E_{j+1}-\E_j$. This level spacing is reported to be Wigner-Dyson like the Gaussian Orthogonal RMT ensemble for non-integrable triangles; and, for pseudo-integrable triangles, tends to  Wigner-Dyson (WD) for small spacings ($s\to0$) and Poisson for large spacings ($s\to\infty$) \cite{Miltenburg1994,Lozej2022,yczkowski1992ClassicalAQ,Cheon:levelpseudointegrable,Bogomolny2021,Gorin_2001,barrierbilliards}.

\begin{figure}
    \centering
    \includegraphics[width=\textwidth]{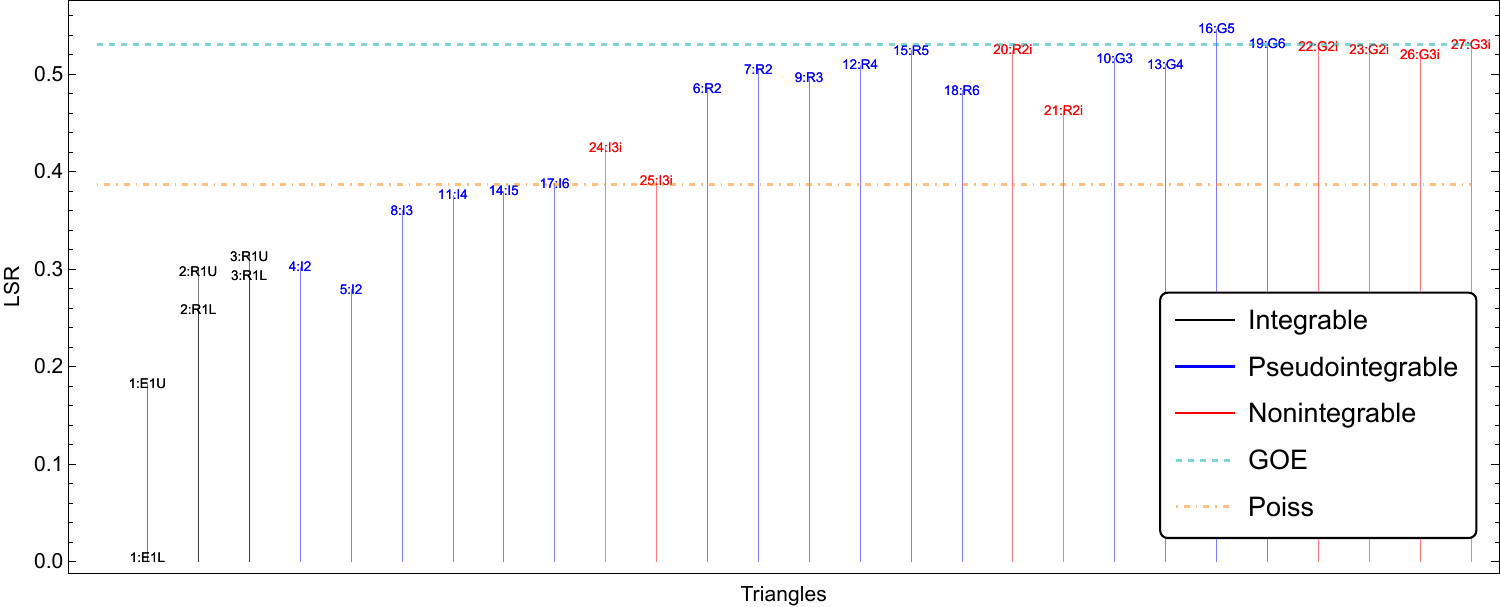}
    \caption{Level spacing ratio (LSR)  for all 27 triangles. The data points are marked by labels of the form $\chi:Tg$. Here, $\kappa$ is the index of the triangle from Table \ref{tab:trig_list},  $T$ is the triangle type  ($E$ = equilateral, $I$ = isosceles, $R$ = right, $G$ = general), and $g$ is the genus of the translation surface of the triangle. For irrational triangles which have infinite genus, we replace $g$ by $2i$ or $3i$ for triangles with 2 and 3 irrational angles, respectively. Sky blue dashed line = LSR for the Gaussian Orthogonal Ensemble (GOE). Orange dot-dashed line =  LSR value for the Poisson distribution.  Integrable and isosceles triangles approach the Poisson result. Generic and right triangles approach the GOE result. We report lower and upper average LSRs for the three integrable triangles by replacing all ill-defined $0/0$ level spacing ratios with either $0$ or $1$.
   }
    \label{fig:LSRDistribution}
\end{figure}

\begin{figure}
    \centering
    \includegraphics[width=\textwidth]{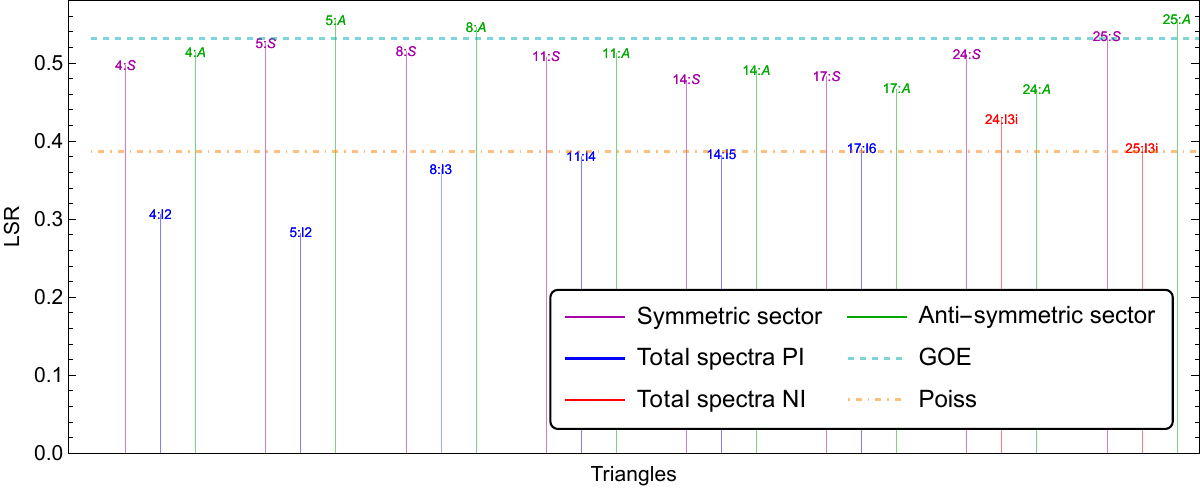}
    \caption{Level spacing ratio (LSR) for isosceles triangles.  Data points marked using the same labels as Fig.~\ref{fig:LSRDistribution}. PI =  pseudo-integrable and NI = non-integrable. 
    $\kappa$:S and $\kappa$:A denote the symmetric and anti-symmetric sectors of the triangle indexed by $\kappa$. The total spectra of isosceles triangles have LSRs closer to Poisson, but their symmetric and anti-symmetric sectors have LSRs closer to  GOE.}
    \label{fig:LSR_iso_sectors}
\end{figure}

Define the average Level Spacing Ratio (LSR) as \cite{Oganesyan2007LSR}
\begin{align}
    \bar r=\frac1{D_0-1}\sum_{n=1}^{D_0-1} \min\kc{r_j,\frac1{r_j}},\quad \text{ where } r_j=\frac{s_j}{s_{j+1}}.
\end{align} 
The minimum means that we take the ratio of the smaller interval to the larger one so that  $r_j \in [0,1]$.  Since the LSR involves consecutive levels, it is independent of the local density of states and unfolding.   Regular level spacing gives $\bar{r}=1$, and larger fluctuations produce a smaller average LSR.  Poisson spectra have $\bar r_{\rm Poisson}\approx 0.3863$ and GOE spectra have $\bar r_{\rm GOE}\approx 0.5307$ 
\cite{Atas:2013gvn}. For models with multiple degenerate energy levels, the ratio $r_j$ can be ill-defined for some $j$ because it is $0/0$.  For such degenerate cases, we assign either the minimum value $0$ or the maximum value $1$ to all the ill-defined ratios, thus computing an upper and lower average LSR. Fig.~\ref{fig:LSRDistribution} shows that the average LSR for our triangular billiards is organized in a  hierarchy:
\begin{itemize}
\item Generic triangles (4 non-integrable and 4 pseudo-integrable), which lack symmetries or fine-tuned angles, have LSRs matching GOE statistics, suggesting long-term chaos.
\item Right angled triangles (6 pseudo-integrable and 2 non-integrable) have LSRs slightly lower than the GOE.
\item Isosceles triangles (6 pseudo-integrable and 2 non-integrable) have LSRs close to the Poisson value.  However, their symmetric and antisymmetric sectors separately approach GOE statistics (Fig.~\ref{fig:LSR_iso_sectors}), suggesting correlations within but not between sectors.
\item The three integrable triangles have the lowest LSRs, suggesting even greater fluctuations in level spacing than expected from Poisson statistics.  These triangles have substantial degeneracies in their spectra, and thus we report lower and upper average LSRs by replacing all ill-defined $0/0$ level spacing ratios by $0$ or $1$ as described above.
\end{itemize}

\subsection{Spectral complexity}

Next, we use the truncated spectra in \eqref{eq:Truncate} to calculate the spectral form factor (SFF) and the spectral complexity (SC) of our triangular billiards. The spectral form factor is 
\begin{align}
    \SFF(t)=\sum_{jk}^D \exp(it(E_j-E_k))=\int dE_1 \, dE_2 \, \rho(E_1) \, \rho(E_2) \, e^{it(E_1-E_2)}\,,
\label{eq:SFF}
\end{align}
where $D$ is the dimension of the truncated Hilbert space and the spectral density is
\begin{align}
    \rho(E)=\sum_j \delta(E-E_j)
\end{align}
with normalisation $\int dE \, \rho(E)=D$. If the spectrum is non-degenerate, the long time limit of the SFF converges to a plateau value $D$ \cite{Cotler:2016fpe}, arising from the  $j=k$ terms in \eqref{eq:SFF}.
The spectral complexity is 
\begin{align}\label{eq:SC}    
C_S(t)=\sum^D_{j\neq k} \kd{\frac{\sin(t(E_j-E_k)/2)}{D(E_j-E_k)/2}}^2
=\int dE_1 \, dE_2 \, \rho(E) \, \rho(E_2) \, \kd{\frac{\sin(t(E_1-E_2)/2)}{D(E_1-E_2)/2}}^2,
\end{align}
For systems with non-degenerate spectra, since $\sin^2(t(E_j-E_k)/2)\leq 1$ in \eqref{eq:SC}, the SC is bounded above by $\kc{\frac2{\Delta E_{\min}}}^2$, with $\Delta E_{\min}$ the minimal level spacing. Since we have rescaled the spectrum to have a width of 1, the minimal level spacing $\Delta E_{\min}$ scales as $1/D$. For systems with degenerate spectra, taking the limit $E_j - E_k \to 0$ in \eqref{eq:SC} reveals that SC will grow as $\frac{d}{D^2} t^2$ at late times, with $d=\sum_{j\neq k}\delta_{E_jE_k}$.   Provided the SFF approaches a late time plateau of $D$, as it does for  non-degenerate spectra, the SFF and SC are simply related by \cite{Iliesiu:2021ari,Erdmenger:2023wjg}
\begin{align}
    C_S(t)=\frac2{D^2}\int_0^tdt'\int_0^{t'} dt'' (\SFF(t'')-D). \label{eq:SFF_SC}
\end{align}

First we work out the result for spectra with Poisson and GOE statistics.  As discussed in \eqref{eq:Truncate} we work in an energy  for the triangle billiards in which the cumulative number of states grows linearly with energy, so that the coarse-grained spectral density is constant
\begin{align}\label{eq:SpectralDensity}
    \bar\rho(E)= d\bar N/dE=D.
\end{align}
It would be useful to know how the SFF and SC behave for theories with the coarse-grained level density \eqref{eq:SpectralDensity} and either no spectral correlations, as expected for an integrable theory, or GOE correlations, as expected for a maximally chaotic theory.  We can study this by taking the expectation value of \eqref{eq:SFF} and \eqref{eq:SC} over Poisson or Gaussian Orthogonal ensembles. Doing so amounts to replacing the $\rho(E_1) \rho(E_2)$ factors by ensemble average, which gives the density-density correlation function. 

If the theory is integrable, we expect that there are no spectral correlations so that the density-density correlation function at large $D$ is
\begin{align}
    \avg{\rho(E_1)\rho(E_2)}
    =\bar\rho(E)\delta(s)
    +\kc{1-\frac1D}\bar\rho(E_1)\bar\rho(E_2) \, ,
\end{align}
where $E=(E_1+E_2)/2,\ s=E_1-E_2$ \cite{Haake2010}. 
Inserting this into \eqref{eq:SFF}, along with the coarse-grained spectral density \eqref{eq:SpectralDensity}, we obtain the SFF
\begin{align}
    \SFF(t)=(D-1) D \sinc^2\frac t2 +D
    \approx 
    \begin{cases}
        4D^2\sinc^2{\frac t2}\,, & t\ll \sqrt{D} \\
        D\,, & \sqrt D\ll t 
    \end{cases}  \, ,
\end{align}
where $\sinc t=\frac{\sin t}{t}$.  This SFF slopes quadratically down a maximum value of $D^2$ until $t \sim \sqrt{D}$ when it crosses over to a plateau value of $D$. For Poisson-distributed spectra with finite $D$, exact degeneracies are measure 0, so we can apply \eqref{eq:SFF_SC} to compute the SC.
Since the SFF is always greater than or equal to its plateau value $D$, the integrand of \eqref{eq:SFF_SC} is always non-negative, which implies eternal growth of the SC at late times. Indeed by performing the double integral over time in \eqref{eq:SFF_SC}  we obtain the spectral complexity
\begin{align}\label{eq:SC_Poisson}
    C_S(t)=4\kc{1-\frac1D} (\Ci(t)+t \Si(t)-\log (t)+\cos (t)-\gamma -1)
    \approx \begin{cases}
        t^2, & t\ll1 \\
        2\pi t, & 1\ll t \, ,
    \end{cases}
\end{align}
with $\Ci(t)=-\int_t^\infty dz\frac{\cos z}{z},\ \Si(t)=\int_0^t dz\frac{\sin z}{z}$, and Euler's constant $\gamma$. Thus, for Poisson distributed spectra, the SC grows linearly at late times.

For random matrix theories drawn from the GOE, the spectrum is correlated as:
\begin{align}\label{TwoPoints}
    \avg{\rho(E_1)\rho(E_2)}
    =\bar\rho(E)\delta(s) +\bar\rho(E_1)\bar\rho(E_2)(1-\hat k(\pi D s)),
\end{align}
where 
\begin{align}
    \hat k(x)=\sinc^2x+\frac1{x}(\Si(x)-\frac\pi2  \sgn x) (\sinc x-\cos x)
\end{align}
is related to the well-known sine kernel in the GOE \cite{Haake2010,Cotler:2016fpe,Liu:2018hlr}. 
Inserting the two-point function into \eqref{eq:SFF}, along with the spectral density \eqref{eq:SpectralDensity}, we can approximate the integral $\iint_0^1 dE_1dE_2 \hat k(\pi D s)e^{its}\approx \int_0^1 dE\int_{-\infty}^{\infty} ds \hat k(\pi D s)e^{its}$ at large $D$, because the kernel $\hat k(\pi D s)$ is localized at $s=0$ with a  width $1/D$ \cite{Liu:2018hlr}. Then the SFF is
\begin{align}
    \SFF(t)
    =&~D^2 \sinc^2\frac t2+D
    +
\begin{cases}
 \frac{t}{\pi}-D-\frac{t}{2\pi} \log \kc{1+\frac t{\pi D}}, & t<2\pi D \\
 D-\frac t\pi \coth ^{-1}\frac{t}{\pi D}, & t\geq 2\pi D \,  \\
\end{cases}\\
\approx &~ \begin{cases}
    4D^2\sinc^2\frac t2\,, & t\ll D^{2/3}\\
    \frac t \pi\,, & D^{2/3}\ll t \ll D \\
    D\,, & D\ll t
\end{cases}.
\end{align}
It exhibits a slope down to a minimum at a ``dip time'' $t\sim D^{2/3}$, and then ramps linearly upward up to a late time plateau of magnitude $D$. 
Since the SFF is smaller than its plateau value $D$ during the ramp, \eqref{eq:SFF_SC} implies that the spectral complexity should grow more slowly after the dip time. Inserting the two-point function into \eqref{eq:SC} and again using the localisation of $\hat k(\pi D s)$, we obtain the spectral complexity 
\begin{align}\label{eq:SC_GOE}
    C_S(t)
    =&~4(\Ci(t)+t \Si(t)-\log (t)+\cos (t)-\gamma -1)+D C_1\kc{\frac t{\pi D}}\\
    \to&~ \begin{cases}
        t^2, & t\ll1 \\
        2\pi t, & 1\ll t\ll D\\
        \frac{2}{3} \pi ^2 D \log \left(\frac{t}{\pi  D}\right),& D\ll t
    \end{cases}
\end{align}
where
\begin{align}
    C_1(u)= ~\frac{\pi^2}{3}
\begin{cases}
    {u \kc{\frac{17 }{12}u^2-4 u-1}+ (1-\frac u2) (u+1)^2 \log (u+1)}, & u<2 \\
    {\log \left(u^2-1\right)-\left(u^2-3\right) u \coth ^{-1}u+ (u-6) u+\frac 43}, & u>2 \\
\end{cases}.
\end{align}
Thus, at late times,  spectral complexity for the GOE exhibits logarithmic growth.

\begin{figure}
    \centering
    \includegraphics[height=0.6\linewidth]{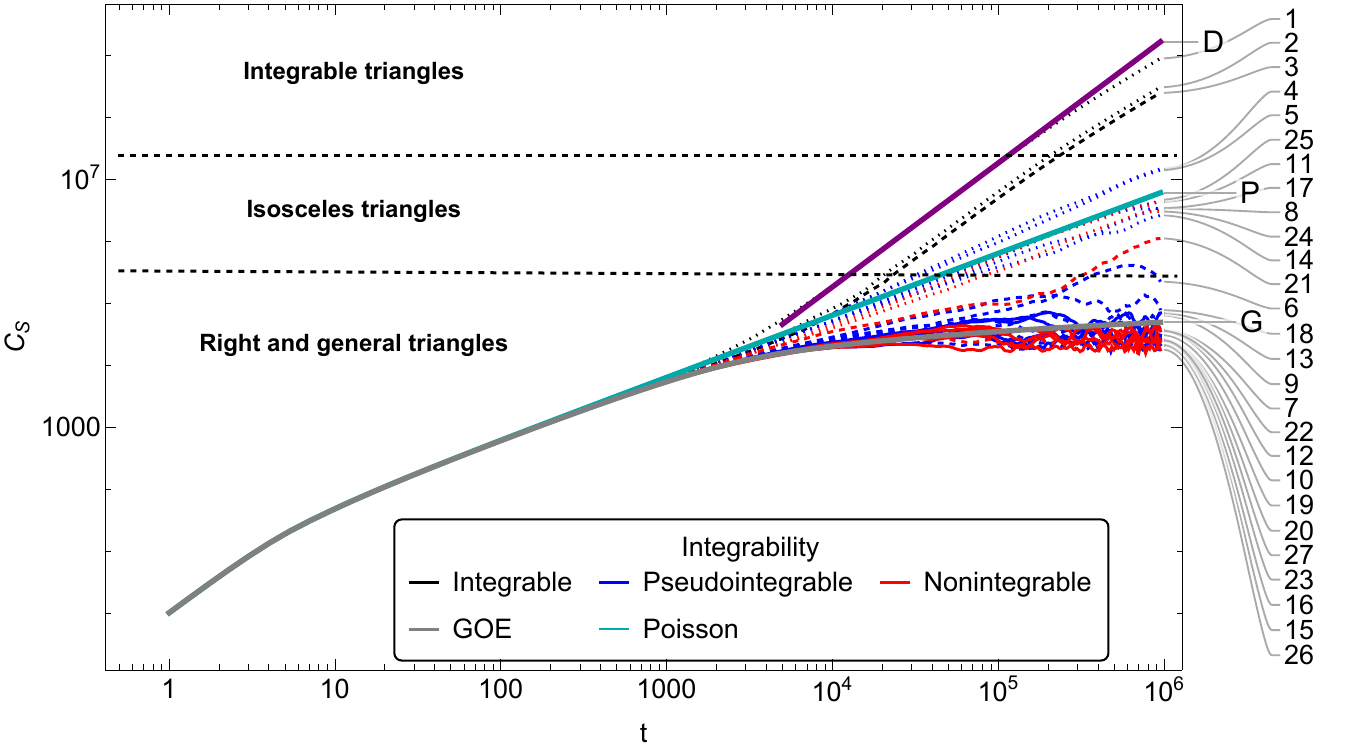}
    \caption{Spectral complexity of all the 27 billiards. The numbers on the right show the $\kappa$ index of the triangles as defined in Table \ref{tab:trig_list}. The labels P, G and D denote our analytical prediction of the spectral complexity growth for models with Poisson (proportional to $t$ at late times), GOE (logarithmic in $t$ at late times) and degenerate spectra (proportional to $t^2$ at late times) with the same degeneracy as that in the equilateral triangle $1$. Dotted lines are for isosceles triangles, dashed lines are for right triangles, and solid lines (except the three denoted P, G and D) are for general triangles. The integrable triangles with degenerate spectra show quadratic late time growth, while the non-integrable triangles match logarithmic late time growth except for the GOE. The isosceles triangles show linear late time growth similar to that expected for a Poisson spectrum.} \label{fig:SCall}

\end{figure}

\begin{figure}
    \centering
    \includegraphics[height=0.6\linewidth]{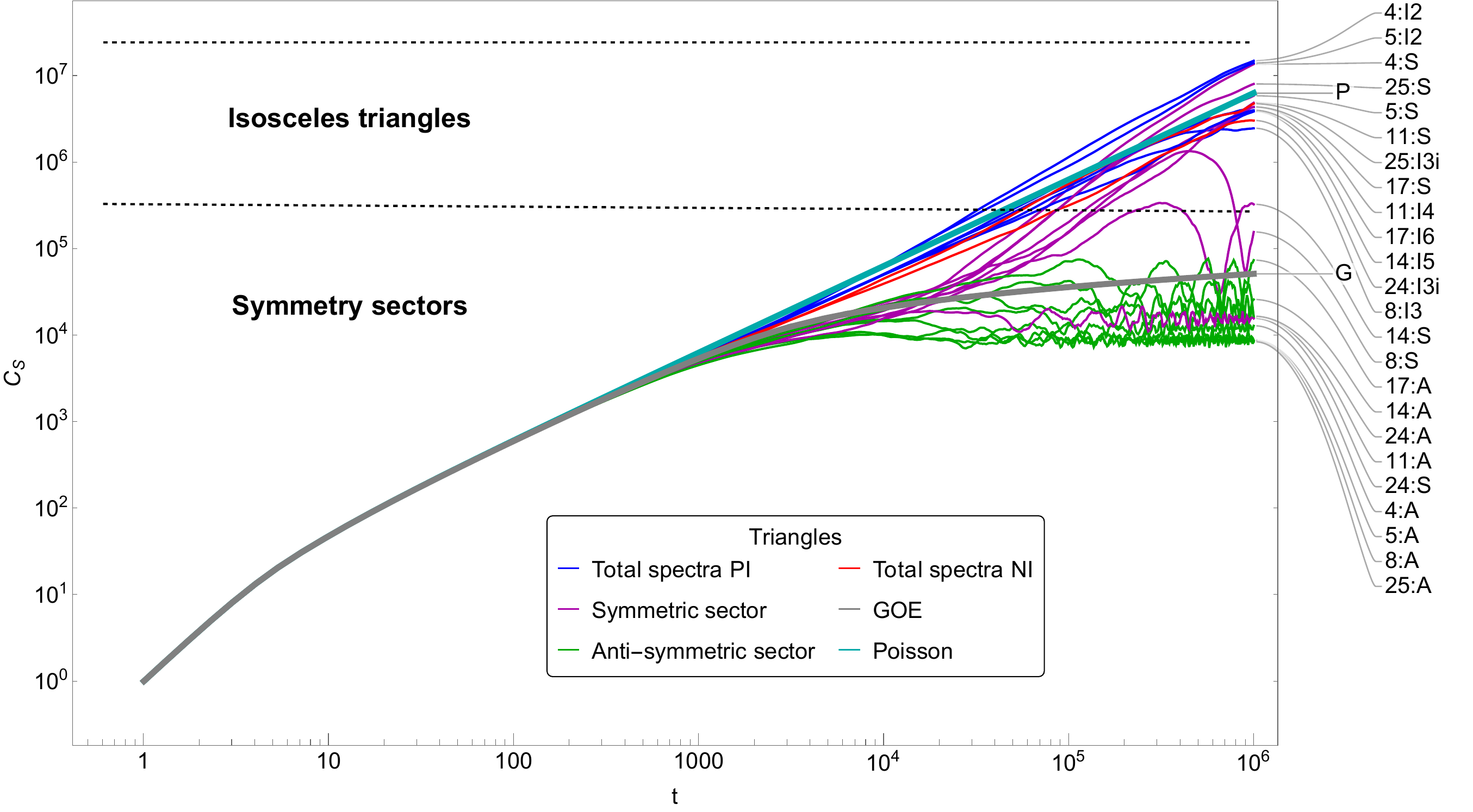}
    \caption{Spectral complexity for isosceles triangles. Labels match those in Fig. \ref{fig:LSR_iso_sectors}. The labels P and G denote the Poissonian and GOE spectral complexity, respectively. The isosceles triangles show linear growth at late times, but their anti-symmetric sectors more closely match the GOE behaviour, while their symmetric sectors also deviate dramatically from linear growth.  Both symmetry sectors show large late time oscillations.
    } \label{fig:SC_iso_sectors}

\end{figure}

The above analytical calculations  show that, for the same spectral density, uncorrelated spectra lead to faster late time growth of SC than for GOE-correlated spectrum  (Fig.~\ref{fig:SCall}.) Note that these expressions are valid in the leading large $D$ limit for rescaled spectra with a bounded range. In this limit, the minimum gap $\Delta E_{\min}$ goes to zero, so the SC bound of $\kc{\frac2{\Delta E_{\min}}}^2$ becomes trivial. 
Fig.~\ref{fig:SCall} shows that the late time growth of spectral complexity for our triangular billiards shows a variety of behaviours:
\begin{itemize}
\item Generic triangles (4 non-integrable and 4 pseudo-integrable), which lack symmetries or fine-tuned angles, show slow late time growth like the GOE.
\item Right angled triangles (6 pseudo-integrable and 2 non-integrable) also show slow late growth of spectral complexity but, on average, reach slightly higher late time values. 
\item Isosceles triangles (6 pseudo-integrable and 2 non-integrable) closely match Poisson late time linear growth of spectral complexity.  However,  Fig.~\ref{fig:SC_iso_sectors}, shows that the antisymmetric sectors of all these theories show slow late time growth comparable to the GOE, while the symmetric sectors typically show a longer period of linear growth before slowing down, and even in some cases decreasing markedly towards the GOE.   Both symmetric and antisymmetric sectors also show large late time oscillations of spectral complexity.   It would be interesting to understand why the SC of these two sectors behaves so differently, although their LSRs are similar (Fig.~\ref{fig:LSR_iso_sectors}).
\item The three integrable triangles have substantial spectral degeneracies which lead to late time quadratic SC growth as expected from the discussion below \eqref{eq:SC}.
\end{itemize}

\section{Spread complexity in triangular billiards}
\label{sec:krylov_results}
\begin{figure}
    \centering
    \includegraphics[width=\textwidth]{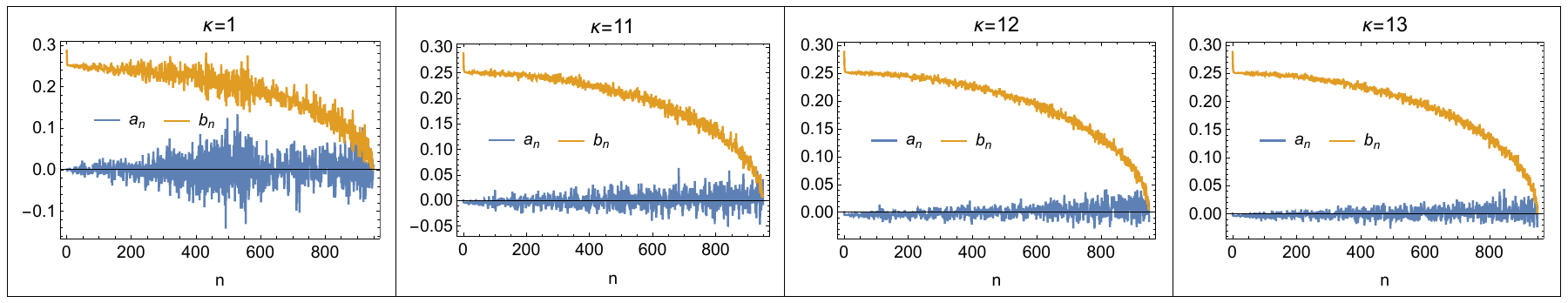}
    \caption{Left to right: Lanczos coefficients $a_n$ and $b_n$ for $\kappa$=1 (integrable, equilateral), $\kappa$=11 (pseudo-integrable, isosceles), $\kappa$=12 (pseudo-integrable, right), $\kappa$=13 (pseudo-integrable, general) triangles. The variance is the highest for the integrable triangle.}
    \label{fig:LanczosCoefficients}
\end{figure}

Next, we will show how the spread complexity and other Krylov space probes, such as the variance of the Lanczos coefficients and the inverse participation ratio of energy eigenstates in the Krylov space, distinguish integrable and chaotic dynamics in triangular billiards \cite{Edwards_1972, Misguich_2016, Rabinovici:2021qqt, Bhattacharjee:2024yxj}. As the Hilbert space of these systems is infinite-dimensional, we first truncate it for operational purposes to the  $D=1000$ lowest energy eigenstates, over which the initial state is uniformly spread 
\begin{align}
    |\psi(0)\rangle=\frac1{\sqrt D}\sum_{j=0}^{D-1}| E_j\rangle \, .
\end{align}
The state undergoes time evolution under the Hamiltonian: $|\psi(t)\rangle=e^{-iHt}|\psi(0)\rangle$. We calculate the Lanczos coefficients $\{a_n,b_n\}$ and build the orthogonal Krylov basis $\{|O_n\rangle\}_{n=0}^{D-1}$ following 
\cite{Viswanath1994,Lanczos:1950zz}:
\begin{align}
	 \ket{O_0} & = \ket{\psi(0)}, \quad b_0=0,\nn\\
	b_n\ket{O_n} & = (H -a_{n-1})\ket{O_{n-1}} - b_{n-1} \ket{O_{n-2}}, \quad 1\leq n \leq K-1,\label{Algorithm}\\
	 a_n & =\bra{O_n}H\ket{O_n},\quad \avg{O_n|O_n}=1.\nn
\end{align}
The algorithm terminates at some $n=K$ when $b_K=0$ giving a  Krylov space of dimension $K\leq D$ (examples in Fig.~\ref{fig:LanczosCoefficients}).  This procedure is equivalent to tridiagonalizing the  Hamiltonian:
 \begin{equation}\label{eq:tridiagonal}
	T_{mn}=\bra{O_m}H\ket{O_n}
	=\begin{pmatrix}
		a_0		&	b_1	&	0	&	\cdots	&	0	\\
		b_1		&	a_1	&	b_2	&	\cdots	&	0	\\
		0		&	b_2	&	a_2	&	\cdots	&	0	\\		\vdots	&\vdots	&\vdots	&   \ddots	&b_{K-1}\\
		0		&	0	&	0	&	b_{K-1}&a_{K-1}
	\end{pmatrix}.
\end{equation}

To quantify how the wavefunction spreads, we measure the support of the time-evolved state has on  Krylov basis vectors:
\begin{align}
    |\phi_n(t)|^2=|\langle O_n|\psi(t)\rangle|^2 \,.
\end{align}
We can characterize this distribution in terms of its moments
\begin{align}
    C^{(\alpha)}_K(t)=\sum_{n=0}^{N-1} n^\alpha |\phi_n(t)|^2. \label{eq:kc_def}
\end{align}
The $\alpha=1$ moment is called the spread complexity \cite{Balasubramanian:2022tpr}, but the higher moments are useful to consider because they amplify features of the dynamics associated to chaos \cite{Balasubramanian:2022tpr,Erdmenger:2023wjg}.

\subsection{Variance of the Lanczos coefficients}
\label{sec:lanczos_results}

It has been observed that higher classical Lyapunov exponents are associated with lower variance in the Lanczos coefficients \cite{Hashimoto:2023swv}.  But in our case, the classical Lyapunov exponents vanish, as they do for all polygonal billiards \cite{Jain_2017}.  On the other hand, the authors of \cite{Hashimoto:2023swv,Balasubramanian:2023kwd} proposed that the variance of the Lanczos coefficients provides a signature discriminating between integrability and chaos. 
We therefore calculate the standard deviations 
    \begin{align}
        \sigma(a_n)=\sqrt{\sum_n x_n^2-\kc{\sum_n x_n}^2},\quad x_n=a_{n+1}-a_n\\
        \sigma(2b_n)=\sqrt{\sum_n y_n^2-\kc{\sum_n y_n}^2},\quad y_n=2b_{n+1}-2b_n.
    \end{align}
We are studying the standard deviations of $\ke{a_n,2b_n}$ because, for a random state initial state and evolution with a random matrix Hamiltons, their variances $a_n$ and $2b_n$ are equal \cite{Balasubramanian:2022dnj,Balasubramanian:2023kwd}.  Note that these definitions of variance differ from those used in \cite{Rabinovici:2021qqt,Hashimoto:2023swv}.

\begin{figure}[hbtp]
   \centering
   \includegraphics[width=0.9\linewidth]{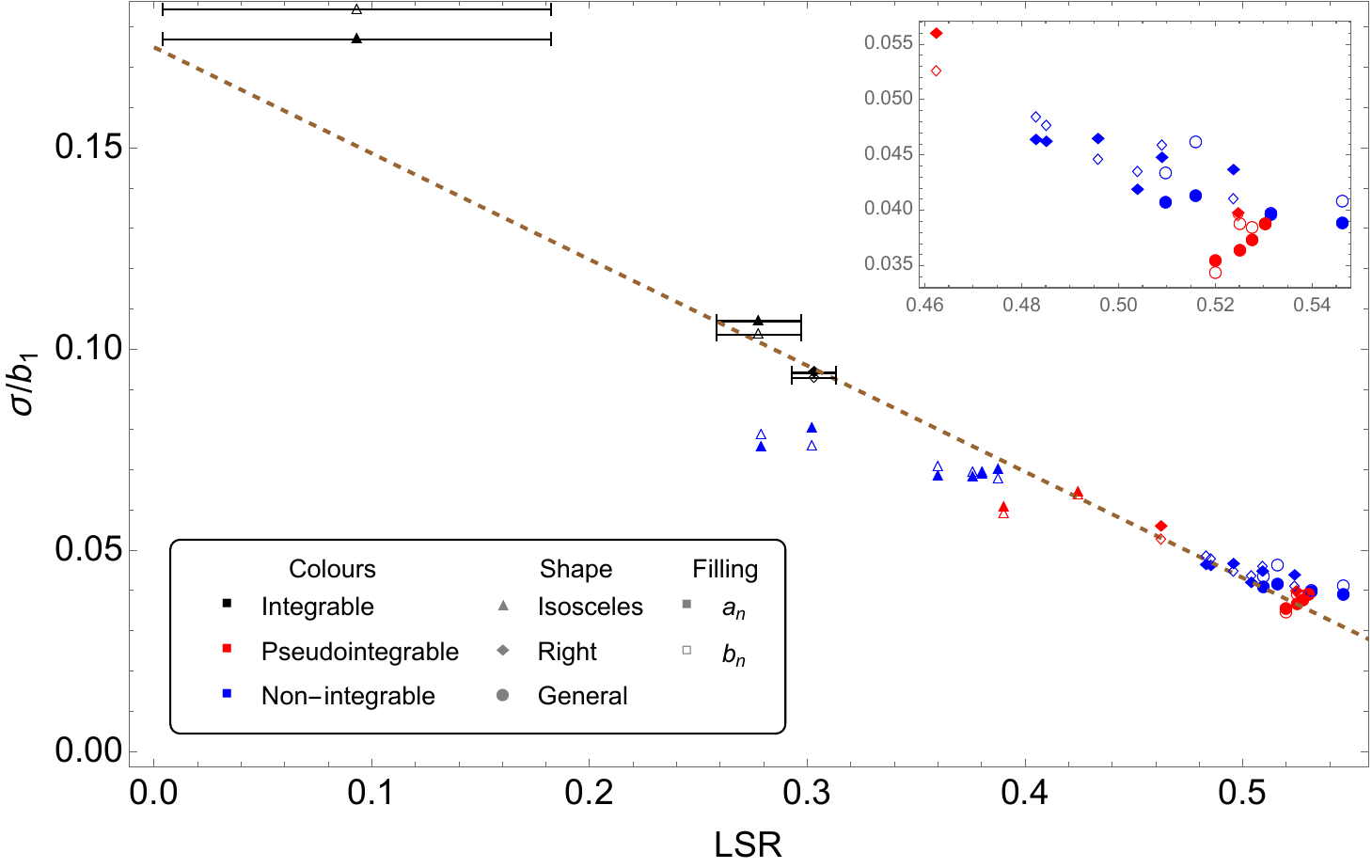}
   \caption{Standard deviations normalized by $b_1$ ($\sigma/b_1$) of $\ke{a_n,2b_n}$ vs. level spacing ratio for all  27 triangle billiards.  For the integrable triangles, the error bars represent the range between the upper and lower average LSRs computed by replacing all ill-defined $0/0$ level spacing ratios by either $0$ or $1$. Small values of level spacing ratio correlate to high values of the standard deviation of the Lanczos coefficients. The dashed brown line is a linear fit of the data points:   using the average of the upper and lower LSRs for integrable triangles, we find a  slope of $-0.262$ and a positive y-axis intercept of 0.174. Inset: Focus on non-isosceles, pseudo-integrable and non-integrable triangles.
   }
   \label{fig:LSR_SV}
\end{figure}

Fig.~\ref{fig:LSR_SV} shows a clear trend. Integrable triangles have by far the largest variance in their Lanczos coefficients, followed by isosceles triangles.  Right triangles come next, with Lanczos variance generally larger than those of the generic pseudo-integrable triangles, which are in turn larger than those of non-integrable triangles (Fig.~\ref{fig:LSR_SV} inset).  Interestingly, the Lanzos variances are anti-correlated with the Level Spacing Ratio (LSR): the higher the LSR, the lower the Lanczos variance (Fig.~\ref{fig:LSR_SV}).  Low LSRs suggest high level repulsion and spectral correlation, perhaps leading to less statistical independence, and hence variance, between the Lanczos coefficients.

\subsection{Localisation of energy eigenstates in Krylov space}
\label{sec:Localisation_results}
\begin{figure}[hbtp]
    \centering
    \includegraphics[width=\textwidth]{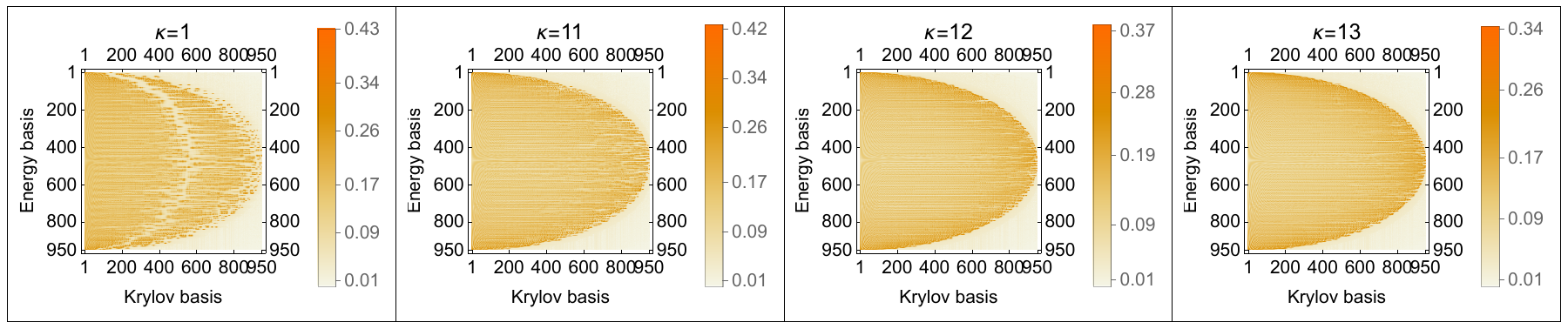}
    \caption{Left to right: Overlap matrix of the energy eigenstates with the Krylov basis, $|\langle E_j|O_n\rangle|$ for $\kappa$=1 (integrable, equilateral), $\kappa$=11 (pseudo-integrable, isosceles), $\kappa$=12 (pseudo-integrable, right), $\kappa$=13 (pseudo-integrable, general) triangles. The localisation of energy basis on the Krylov basis vectors is the highest for the integrable triangles (see text), but this concentration is not readily visible because nearby energy eigenstates can localize onto different Krylov states. }
    \label{fig:loc_matrix}
\end{figure}

Next, we consider the overlap $\abs{\langle O_n|E_j\rangle}$ between the energy basis and the Krylov basis (Fig.~\ref{fig:loc_matrix}). 
The overlap $\abs{\langle O_n|E_j\rangle}$ is approximately zero for energies $\abs{E_j}>2b_n$  \cite{Erdmenger:2023wjg}.  This concentration can be understood by noting that, as shown in Eq.~39 of \cite{Balasubramanian:2023kwd},
\begin{equation}
\sum_{E' \in [E,E+dE]} | \langle K_m | E' \rangle |^2 \approx \frac{\Theta(4b_m^2-(E-a_m)^2)}{\pi\sqrt{4 b_m^2-(E-a_m)^2}} \, dE
\end{equation}
where $dE$ is a small interval of energies.   Since $b_m$ is on average a decreasing function of $m$ and $a_m$ is zero on average  (Fig.~\ref{fig:LanczosCoefficients}), $\langle O_n|E_j\rangle$ tends to vanish for $\abs{E_j}>2b_n$.  Thus,  low energy and high energy eigenstates tend to be located in the Krylov basis with small $n$ (Fig.~\ref{fig:loc_matrix}). Note that this is a general phenomenon that depends on the coarse-grained spectral density and is independent of integrability or chaos in the dynamics.

\begin{figure}[hbtp]
    \centering
    \includegraphics[width=\textwidth]{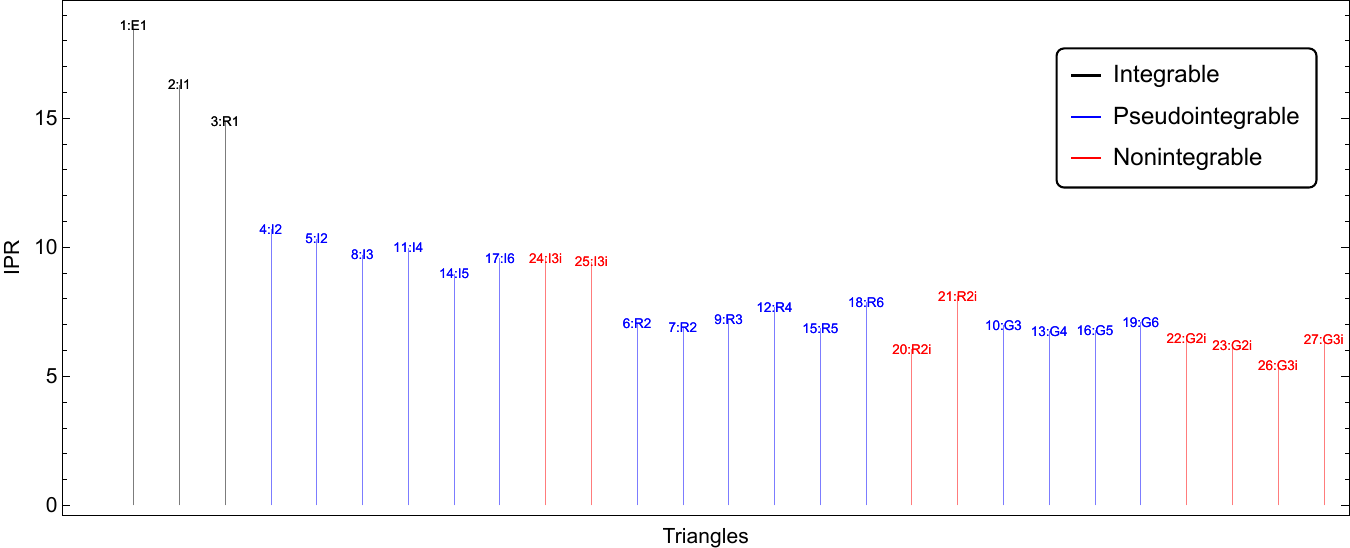}
    \caption{Inverse participation ratio for all  27 triangles.  Data points are marked by labels of the form $\chi:Tg$. Here, $\kappa$ is the index of the triangle from Table \ref{tab:trig_list},  $T$ is the triangle type  ($E$ = equilateral, $I$ = isosceles, $R$ = right, $G$ = general), and $g$ is the genus of the translation surface of the triangle. For irrational triangles which have infinite genus, we replace $g$ by $2i$ or $3i$ for triangles with 2 and 3 irrational angles, respectively. IPR is highest for integrable triangles, followed by isosceles, right and generic triangles.}
    \label{fig:IPR_diff_type}
\end{figure}

For the integrable triangles, the overlap $\abs{\langle O_n|E_j\rangle}$ for a given $j$ concentrates at some $n$ although this is not readily visible in Fig.~\ref{fig:loc_matrix} because nearest neighbour energy eigenstates can localise onto different Krylov basis vectors. 
The authors of \cite{Dymarsky:2019quantum,Rabinovici:2021qqt,Bhattacharjee:2024yxj} explain this sort of phenomenon as a form of Anderson localisation in the Krylov basis, that can be traced to disorder in the Lanczos coefficients of integrable triangles (see related comments in \cite{Balasubramanian:2023kwd}).  
To quantify this localisation, we study the Inverse Participation Ratio (IPR) of energy eigenstates in the Krylov basis:  
\begin{equation}
    \text{IPR}_j=\sum_{n=1}^{K} |\langle O_n|E_j\rangle|^4.
\end{equation}
IPR$_j$ quantifies how localized an energy eigenstate is in the Krylov basis.
If an eigenstate is completely localized,  $|E_j\rangle=|O_n\rangle$, so that IPR$_j = 1 $.
Conversely, if a state is entirely delocalized,  $|E_n\rangle=\frac{1}{\sqrt{K}}\sum_{j=1}^{K}|O_j\rangle$, and IPR$_j = 1/K$.
The higher the IPR, the more pronounced the localisation.
The IPR for the entire system is defined by summing over all the energy eigenstates:
\begin{equation}\label{eq:IPR_Krylov}
    \text{IPR}=\sum_{j=1}^{D} \text{IPR}_j=\sum_{j=1}^{D}\sum_{n=1}^{K} |\langle O_n|E_j\rangle|^4. 
\end{equation}
Fig.~\ref{fig:IPR_diff_type} shows a clear hierarchy.  The integrable triangles show the greatest localisation of energy eigenstates, followed by isosceles triangles, right triangles and generic pseudo-integrable and non-integrable triangles.

\subsection{Spread complexity}
\label{sec:spread_moments_results}
\begin{figure}
     \centering
     \begin{subfigure}[b]{0.49\textwidth}
         \centering
         \includegraphics[width=0.95\textwidth]{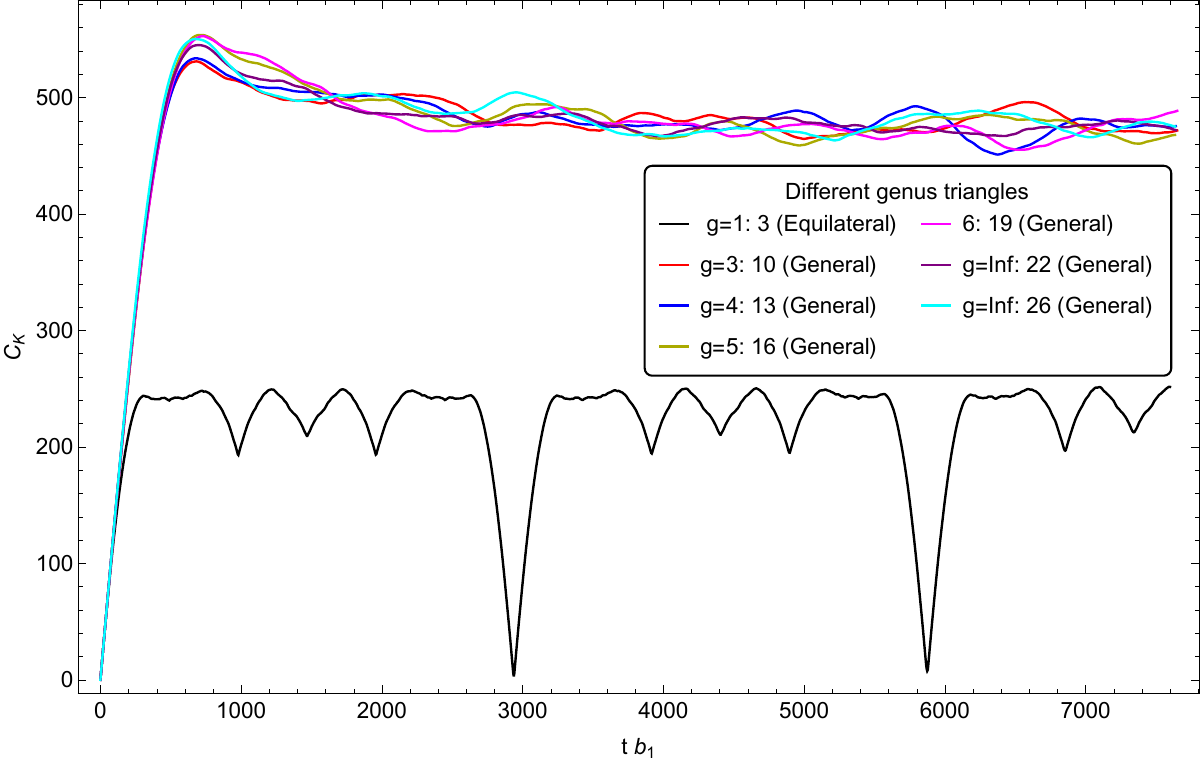}
         \caption{Spread complexity ($\alpha=1$ in Eq.\eqref{eq:kc_def}) for different genus of triangular billiards.\\}
         \label{fig:kry_com_diff_gen}
     \end{subfigure}
     \hfill
     \begin{subfigure}[b]{0.49\textwidth}
         \centering
         \includegraphics[width=\textwidth]{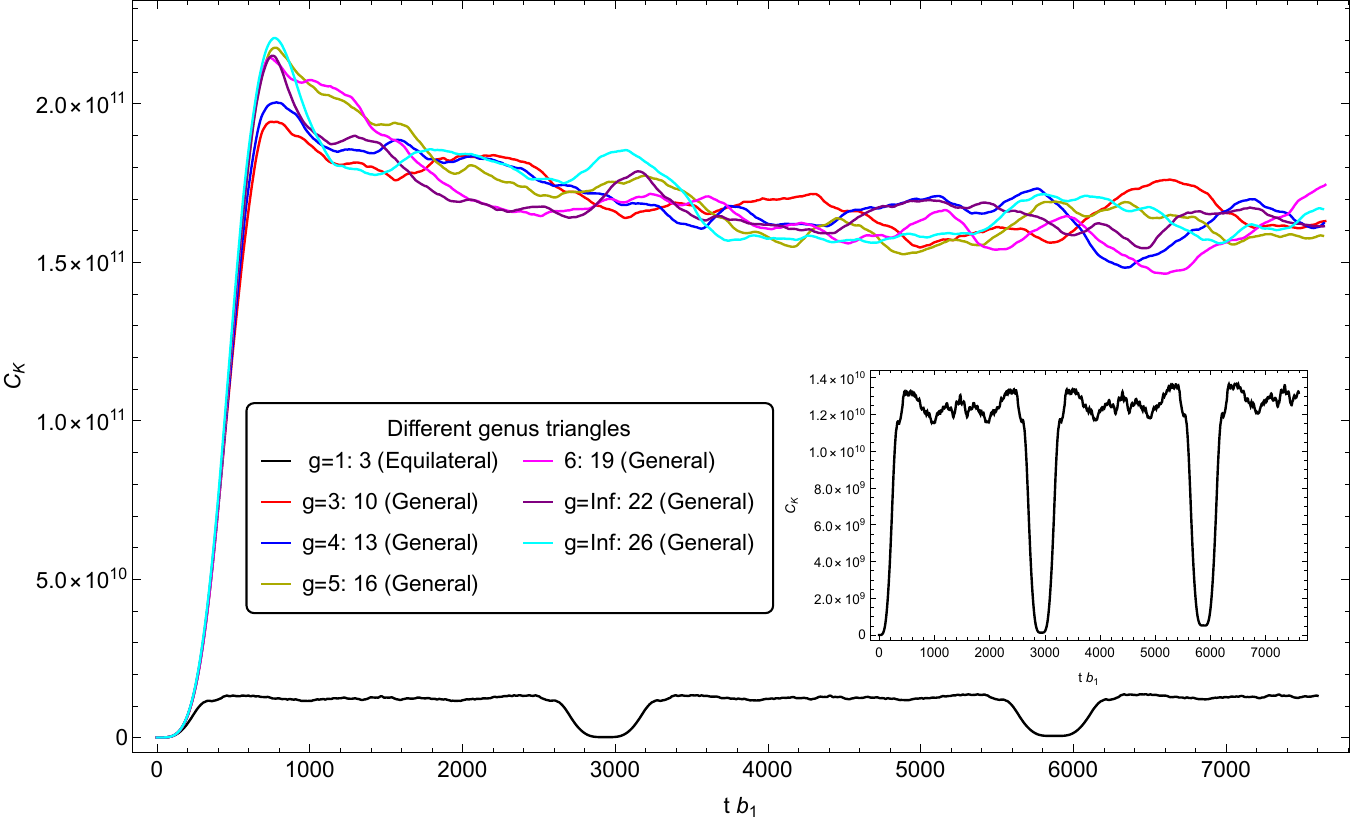}
         \caption{$4$th moment of spread complexity ($\alpha=4$ in Eq.\eqref{eq:kc_def}) for different genus of triangular billiards.}
         \label{fig:kry_com_diff_gen_4th_moment}
     \end{subfigure}
        \caption{Spread complexity and its fourth moment for triangles corresponding to different genera, i.e., $\kappa$=1 (genus one equilateral), $\kappa$=10 (genus three general), $\kappa$=13 (genus four general), $\kappa$=16 (genus five general), and $\kappa$=19 (genus six general) triangle. The integrable triangle corresponding to genus one has oscillatory complexity, whereas the pseudo-integrable and non-integrable general triangles show the characteristic behaviour or growth-peak-decay and saturation.}
        \label{fig:kc_diff_gen}
\end{figure}

Finally we compute spread complexity for our triangular billiards (Figs.~\ref{fig:kry_com_diff_gen}, \ref{fig:kry_com_diff_gen_4th_moment}).  There is a striking difference between the integrable triangles and all the rest: the spread complexity rises to a plateau, exhibits some oscillations and then drops dramatically to zero, indicating the start of a recurrence.  This is as expected: in quantum systems with finite dimension $K$, Poincar\'e recurrences are expected to happen at timescale $e^K$ for non-integrable systems and timescale $K$ for integrable systems \cite{PhysRev.107.337}.   Indeed, the spread complexity of the integrable triangle billiards drops to zero periodically, suggesting that the system has returned to its initial state at these Poincar\'e  quantum recurrence times \cite{PhysRev.107.337}.  The recurrences arise from commensurability of the integrable triangle spectra \cite{Hashimoto:2023swv, Doncheski_2002, Jung1980, Damle2010, RICHENS1981495}, and the lower plateau value occurs because of the higher variance of the Lanczos coefficients \cite{Rabinovici:2021qqt}.  By contrast, the general pseudo-integrable and non-integrable triangles all show spread complexity characteristic of chaotic dynamics, namely an initial rise to a peak, followed by a slope downward to a plateau \cite{Balasubramanian:2022tpr,Balasubramanian:2023kwd}.

\begin{figure}
     \centering
     \begin{subfigure}[b]{0.49\textwidth}
         \centering
         \includegraphics[width=0.95\textwidth]{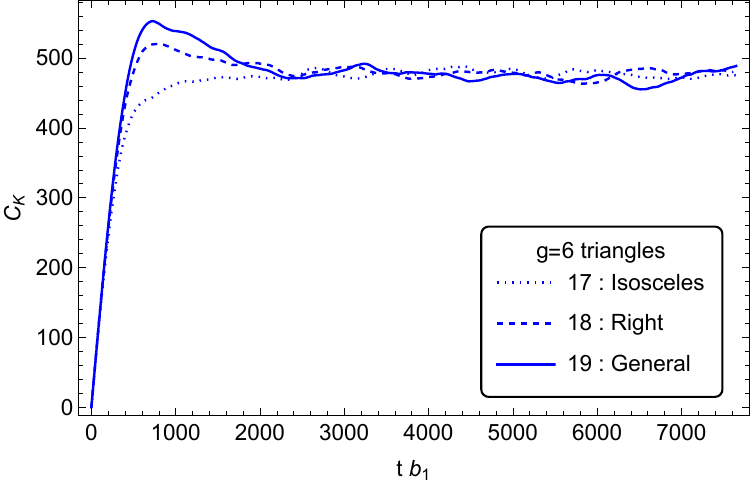}
         \caption{Spread complexity ($\alpha=1$ in Eq.\eqref{eq:kc_def}) for different types of triangular billiards}
         \label{fig:kry_com_diff_type}
     \end{subfigure}
     \hfill
     \begin{subfigure}[b]{0.49\textwidth}
         \centering
         \includegraphics[width=\textwidth]{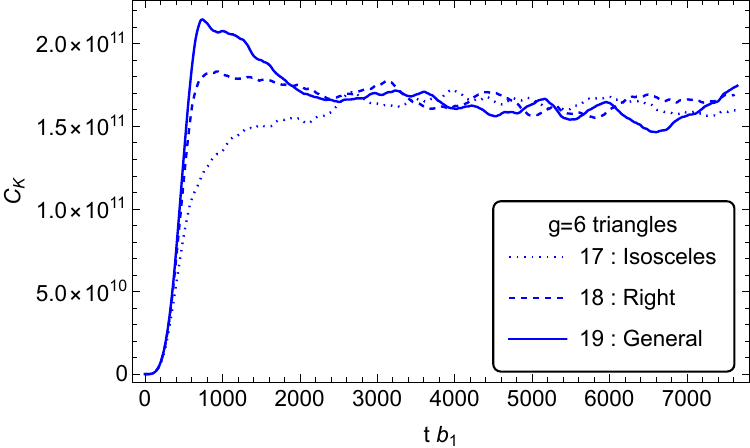}
         \caption{$4$th moment of spread complexity ($\alpha=4$ in Eq.\eqref{eq:kc_def}) for different types of triangular billiards}
         \label{fig:kry_com_diff_type_4th_moment}
     \end{subfigure}
        \caption{Spread complexity and its fourth moment for triangles with genus six translation surfaces: $\kappa$=17 (isosceles), $\kappa$=18 (right) and $\kappa$=19 (general). The height of the complexity peak is reduced by the presence of symmetries. This difference becomes more pronounced for the higher moments of complexity.}
        \label{fig:kc_diff_type}
\end{figure}

\begin{figure}
     \centering
     \begin{subfigure}[b]{0.49\textwidth}
         \centering
         \includegraphics[width=\textwidth]{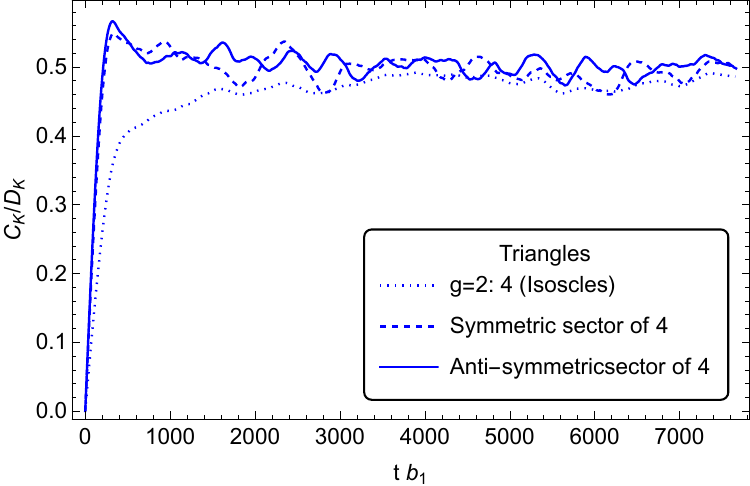}
         \caption{Spread complexity ($\alpha=1$ in Eq.\eqref{eq:kc_def}) for the symmetric and anti-symmetric sectors of isosceles triangular billiard}
         \label{fig:kry_com_diff_sec}
     \end{subfigure}
     \hfill
     \begin{subfigure}[b]{0.49\textwidth}
         \centering
         \includegraphics[width=\textwidth]{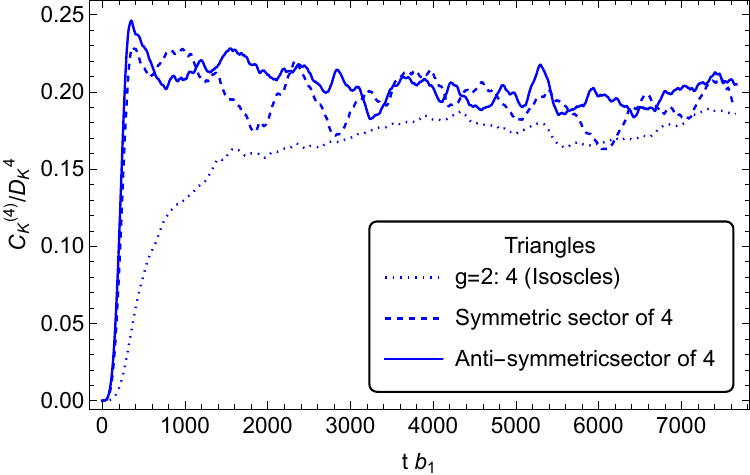}
         \caption{$4$th moment of Spread complexity ($\alpha=4$ in Eq.\eqref{eq:kc_def}) for the symmetric and anti-symmetric sectors of isosceles triangular billiard}
         \label{fig:kry_com_diff_sec_4th_moment}
     \end{subfigure}
        \caption{Spread complexity and its fourth moment the symmetric and anti-symmetric sectors of isosceles triangular billiards with genus two translation surfaces and $\kappa$=4. The individual symmetry sectors display a complexity peak characteristic of chaotic systems, while the complete dynamics does not show this feature.}
        \label{fig:kc_diff_sec}
\end{figure}

Symmetry reduces or eliminates the characteristic feature of chaos -- the spread complexity peak \cite{Balasubramanian:2022tpr,Hashimoto:2023swv,Balasubramanian:2023kwd}: this peak is absent for isosceles triangles and reduced for right triangles, both of which have special symmetries (Fig.~\ref{fig:kc_diff_type}).  The even and odd symmetry sectors of the the isosceles triangles separately show spread complexity complexity growth resembling chaotic theories (rise to a peak, followed by slope down to a plateau), but the complexity growth in the combined dynamics lacks these features (Fig.~\ref{fig:kc_diff_sec}).

\section{Summary and outlook}
\label{sec:Discussion}

In this paper, we surveyed the quantum dynamics of twenty-seven triangular billiards, which can display a variety of dynamics -- integrable, pseudointegrable and non-integrable -- controlled by the internal angles of the triangles.   We characterized these theories spectrally in terms of their level spacing ratios and spectral complexity and then in terms of the variance in their Lanczos coefficients, localisation of energy eigenstates in the Krylov basis, and dynamical growth of spread complexity. 
We find that:
\begin{itemize}
\item {\bf Integrable triangles} have the lowest level spacing ratios (LSRs) and hence the highest level spacing fluctuations.  Degeneracies in their spectrum lead to quadratic late time spectral complexity growth, even faster than the linear growth expected for Poisson-distributed spectra.  Their Lanczos coefficients show the largest variances, and partly as a consequence, their energy eigenstates are strongly localized in the Krylov basis. Finally,  spread complexity for the integrable billiards does not show the complexity peak characteristic of chaos, and also displays marked Poincare recurrences.
\item {\bf Isosceles triangles} from either the pseudo-integrable or the non-nonintegrable classes display LSRs and linearly growing spread complexity growth close to that expected for theories with Poisson spectra. Their Lanczos spectra have variances that are smaller than the integrable triangles, but substantially larger than triangles that lack symmetries.  Likewise, their energy eigenstates are less localized in the Krylov chain than the integrable triangles but more so than generic triangles.  The spread complexity for isosceles triangles does not show the marked recurrences seen for integrable triangles but lacks the peak that is characteristic of chaos.
\item The {\bf symmetric} and {\bf anti-symmetric} sectors of isosceles triangles, from either the pseudo-integrable or the non-nonintegrable classes, separately display LSRs close to the Gaussian Orthogonal Ensemble (GOE) value, but the spectral complexity shows substantial oscillations before approaching the  GOE result, with marked differences between the symmetry sectors.   Similarly, the spread complexity of the two symmetry sectors separately shows a rise to a peak followed by a slope down to a plateau, even though the combined system does not display this characteristic of chaos.
\item {\bf Right triangles} of the pseudo-integrable and non-integrable classes have spectral features that are harder to discriminate from the generic triangle, but the LSRs are, on average, a little lower than the expected GOE value, although significantly higher than the Poisson value.  Consistently, on average, their spectral complexity growth is slightly faster, their Lanczos variance slightly higher, and their energy eigenstate localisation slightly greater than for generic triangles.  The spread complexity growth shows characteristic signs of chaos and does not differ from that of generic triangles at our numerical resolution.
\item {\bf Generic pseudo-integrable triangles} have level spacing ratios and spectral complexity growth close to the predicted behaviour of random matrix theories drawn from the GOE.   The variances of the Lanczos coefficients are low, but slightly higher than for generic non-integrable triangles.  Likewise, their energy eigenstates are less localized than most of the kinds of triangles we studied, but perhaps a bit more so than the general non-integrable triangles (although this requires a more precise numerical study).  Finally, their spread complexity growth closely matches the expectation for chaotic theories. 
\item {\bf Generic non-integrable triangles}  have level spacing ratios and spectral complexity growth that match the expectation for GOE theories. They have the lowest Lanczos coefficient variance, energy eigenstates that are the most delocalized in the Krylov basis, and spread complexity growth that is precisely as expected for chaotic theories.
\end{itemize}

It would be useful to conduct higher precision numerical analyses to better discriminate the pseudo-integrable and non-integrable theories.  Notably, all the triangle billiards have vanishing classical Lyapunov exponent, and chaotic behaviour arises from the pattern of splitting of beams of trajectories at the corners or, equivalently, the topology of the translation surface.  In terms of the latter, there is a dramatic difference between the topology of the finite genus pseudo-integrable and infinite genus non-integrable translation surfaces.  Where does this difference appear in the quantum dynamics?  Possibly a path integral perspective would help, by making explicit trajectories that traverse different handles of the translation surface, and the corresponding contributions to quantum time evolution.  It is also possible that these sorts of effects manifest themselves in patterns of ``scarring'', i.e., localisation of quantum wavefunctions around clusters of classical paths \cite{Heller:stadiumscar,Bogomolny2021}.  

More broadly, the methods for quantifying chaos described above should be extended to quantum field theories, especially theories dual to gravity, in which simple initial states can rapidly form complex final states with macroscopic descriptions as black holes.  It might be helpful to start with simplified models such as the chiral CFTs that are dual to the near-horizon physics of extremal black holes of string theory \cite{Balasubramanian:2003kq,Balasubramanian:2009bg}, or with an appropriately simplified sector of the interacting matrix model dual to M theory \cite{Banks:1996vh,Balasubramanian:1997kd,Polchinski:1999br}. Alternatively, a further step in this direction is to extend the methods discussed here to spin-chain models motivated by holography \cite{Basteiro:2022zur,Basteiro:2022xvu,Basteiro:2024cuh}.

\section*{Acknowledgements}
The authors would like to thank Souvik Banerjee and Yiyang Jia for useful discussions and comments. R.N.D. would like to thank Sudhir Ranjan Jain for the training received during his earlier work with him, which has been helpful for this research. R.N.D., J.E.~and Z.Y.X.~are supported by Germany's Excellence Strategy through the W\"urzburg‐Dresden Cluster of Excellence on Complexity and Topology in Quantum Matter ‐ ct.qmat (EXC 2147, project‐id 390858490),
and  by the Deutsche Forschungsgemeinschaft (DFG) 
through the Collaborative Research centre \enquote{ToCoTronics}, Project-ID 258499086—SFB 1170. R.N.D.~further acknowledges the support by the Deutscher Akademischer Austauschdienst (DAAD, German Academic Exchange Service) through the funding programme, \enquote{Research Grants - Doctoral Programmes in Germany, 2021/22 (57552340)}. ZYX also acknowledges support from the National Natural Science Foundation of China under Grant No.~12075298.
VB was suppported in part by the DOE through DE-SC0013528 and QuantISED grant DE-SC0020360,.  VB also thanks the Aspen Center for Physics (NSF grant PHY-2210452) and the Santa Fe Institute for hospitality as this paper was completed.

\bibliographystyle{JHEP}
\bibliography{biblio.bib}

\end{document}